\documentclass[12pt]{amsart}

\usepackage{amssymb}

\makeatletter
\def\eqalign#1{\null\vcenter{\def\\{\cr}\openup\jot\m@th
  \ialign{\strut$\displaystyle{##}$\hfil&$\displaystyle{{}##}$\hfil
      \crcr#1\crcr}}\,}
\makeatother

\newcommand{\be}{\begin{equation}} 
\newcommand{\ee}{\end{equation}}
\newcommand{\beq}{\begin{eqnarray}}
  \newcommand{\eeq}{\end{eqnarray}}
\newcommand{\bt}{\beta}
\newcommand{\bl}{\begin{lemma}}
\newcommand{\el}{\end{lemma}}
\newcommand{\bm}{\begin{pmatrix}}
\renewcommand{\em}{\end{pmatrix}}
\newcommand{\bml}{\begin{multline}}
\newcommand{\eml}{\end{multline}}
\newcommand{\ba}{\begin{array}}
\newcommand{\ea}{\end{array}}

\newcommand{\la}{\label}
\newcommand{\ci}{\cite}

\newcommand{\de}{\delta}
\newcommand{\De}{\Delta}
\newcommand{\al}{\alpha}

\newcommand{\si}{\sigma}

\newcommand{\Om}{\Omega}
\newcommand{\lb}{\lambda}
\newcommand{\ze}{\zeta}

\renewcommand{\th}{\theta}

\newcommand{\bi}{\bibitem}

\newfont{\msbm}{msbm10 scaled\magstep1}%blackboardbold
\newfont{\msbms}{msbm7 scaled\magstep1} %blackboardbold   subscript

\newcommand{\bbr}{\mbox{$\mbox{\msbm R}$}}

\newcommand{\bbc}{\mbox{$\mbox{\msbm C}$}}

\newtheorem{theorem}{Theorem}[section]
\newtheorem{lemma}[theorem]{Lemma}

\theoremstyle{definition}

\theoremstyle{remark}

\numberwithin{equation}{section}

\newcommand{\Ai}{{\mathrm{Ai}}\,}

\begin{document}
\def\wt{\widetilde}
%\hfill {\small February 28, 2008}
\title{Aspects of Toeplitz determinants} 
\author{I. Krasovsky}
\address{Department of Mathematical Sciences,
Brunel University West London,
Uxbridge UB8 3PH, United Kingdom}
%
%\thanks{The author was supported in part by EPSRC Grant EP/E022928/1.}

%\date{}

\begin{abstract}
We review the asymptotic behavior of a class of Toeplitz (as well as related Hankel and 
Toeplitz + Hankel)
determinants which arise in integrable models and other contexts. We discuss Szeg\H o,
Fisher-Hartwig asymptotics, and how a transition between them is related to 
the Painlev\'e V equation. 
Certain Toeplitz and Hankel determinants reduce, in certain double-scaling limits, to Fredholm 
determinants which appear in the theory of group representations, in
random matrices, random permutations and partitions. The connection to Toeplitz determinants 
helps to evaluate the asymptotics of related Fredholm determinants in situations of interest, and we 
review the corresponding results.
\end{abstract}

\maketitle

\section{Introduction}
Let $f(z)$ be a function integrable over the unit circle $C$ with Fourier coefficients
\[
f_j={1\over 2\pi}\int_0^{2\pi}f(e^{i\theta})e^{-i j\theta}d\theta,\qquad j=0,\pm1,\pm2,\dots
\]
Then the $n$-dimensional Toeplitz determinant of a Toeplitz matrix with symbol $f(z)$ is given by
\be\la{TD}
D_n(f)=\det(f_{j-k})_{j,k=0}^{n-1}.
\ee
Substituting here the expressions for the Fourier coefficients,
and using formulae for Vandermonde determinants, one obtains 
another useful representation:
\be\la{Dint}
D_n(f)={1\over (2\pi)^n n!}
\int_0^{2\pi}\cdots\int_0^{2\pi}\prod_{1\le j<k\le
 n}|e^{i\th_j}-e^{i\th_k}|^2\prod_{j=1}^{n} f(e^{i\th_j}){d \th_j}.
\ee

Toeplitz determinants are closely related to the polynomials orthogonal with weight $f(z)$ 
on the unit circle. Namely, if $D_k(f)\neq 0$, $k=k_0,k_0+1\dots$,
for some $k_0\ge 0$, then the polynomials 
$\phi_k(z)=\chi_k z^k+\cdots$, $\widehat\phi_k(z)=\chi_k z^{k}+\cdots$ 
of degree $k$, $k=k_0,k_0+1,\dots$, satisfying 
\begin{multline}\la{op}
{1\over 2\pi}\int_0^{2\pi}\phi_k(z)z^{-j}f(z)d\theta=\chi_k^{-1}\de_{jk},\qquad
{1\over 2\pi}\int_0^{2\pi}\widehat\phi_k(z^{-1})z^j f(z)d\theta=
\chi_k^{-1}\de_{jk},\\
z=e^{i\theta},\qquad j=0,1,\dots,k,
\end{multline}
exist and $\chi_k=\sqrt{D_k/D_{k+1}}$,
where, by convention, $D_0\equiv 1$.
We see from (\ref{Dint}) that if $f(z)$ is positive on $C$, we have $D_n(f)>0$ for all $n$,
and therefore in this case we can set $k_0=0$.

A Toeplitz determinant can be represented as a Fredholm determinant of an integral operator
acting on $L^2(C)$ which belongs to the special class of so-called {\it integrable operators} \ci{D}.
It can also be written in a different way in terms of a Fredholm determinant of an operator 
now acting on $\ell^2(n,n+1,\dots)$ \ci{GC,BOk,Boe}. 
(Note that the symbols $f(z)$ considered in \ci{GC,BOk,Boe} are assumed to be sufficiently 
smooth.) 
Another useful property of many $D_n(f)$'s encountered in applications
is the existence of simple differential identities
relating the determinant to orthogonal polynomials evaluated at a few special points.
The precise form of such identities depends on the given $f(z)$. 

It turns out that the above properties play a key role in making Toeplitz determinants amenable 
to a detailed asymptotic analysis, in particular, by Riemann-Hilbert-Problem methods. 

In this paper, we will review some asymptotic results on $D_n(f)$ (and related Hankel,
Teoplitz+Hankel, and Fredholm determinants) and briefly 
mention their applications in integrable models, random matrices, random permutations, group
representation theory, and also in various conjectures on Riemann's $\zeta$ and 
Dirichlet's $L$-functions.
This review is based, to a large extent, on the recent work of the author 
with T. Claeys, P. Deift, A. Its, and J. Vasilevska. 
For other aspects of Toeplitz determinants not mentioned here, the reader is referred to
\ci{BGbook,BSKbook,BSbook,GS,DO,DJ,DK} for properties of Toeplitz matrices and determinants, 
to \ci{BSKbook,Wcontin1,Wcontin2,Wcontin3,LaSa,G,Sobolev,BW,L,LS,Wblock,IMM,IJK,BEblock} 
for generalizations to the continuous, higher-dimensional, and the 
block-Toeplitz cases (with relations to stationary determinantal processes, integrable
models, and entanglement entropy), and to \ci{Kmult,AMV,Bertola} for connections with multiple
orthogonal polynomials.

The paper consists of three parts: in Section 2, the simplest asymptotics with $f(z)$
fixed and $n\to\infty$ are considered; in Section 3, the symbol $f(z)$
is allowed to depend on $n$ in ways which describe a transition between
different asymptotic regimes arising in Section 2; in Section 4, the symbol $f(z)$ also depends on $n$,
but in such a way that in the limit $n\to\infty$, Toeplitz determinants turn into
certain Fredholm determinants which are important for random matrices and random permutations.
Following standard practice, we refer to the large $n$ asymptotics of Sections 3 and 4 as 
double-scaling limits.

\section{Asymptotics for a fixed symbol}
We assume in this section that $f(z)$ does not depend on the size of the determinant $n$.
We are interested in the asymptotics of $D_n(f)$ as $n\to\infty$.

The following result is basic.
%In the case when $f(z)$ is sufficiently smooth on the unit circle $C$, there holds the following

\begin{theorem}[Strong Szeg\H o limit theorem]\label{szegothm}
Let $f(z)$ be non-zero on $C$, $\ln f(z)\in L^1(C)$, and suppose that the sum
\be\label{Sf}
S(f)=\sum_{k=-\infty}^{\infty}|k||(\ln f)_k|^2,\qquad 
(\ln f)_k=\frac{1}{2\pi}\int_{0}^{2\pi}\ln f(e^{i\theta})e^{-ik\theta}d\theta,
\ee
converges. Then
\be\label{Sz}
\ln D_n(t)=n(\ln f)_0
+\sum_{k=1}^{\infty}k(\ln f)_k (\ln f)_{-k}+o(1),
\qquad \mbox{ as $n\to\infty$.}
\ee
\end{theorem}
The theorem was initially proved by Szeg\H o \ci{Sz1,Sz2} 
(the leading term in 1915, the next in 1952)
under stronger conditions on $f(z)$. The conditions 
were then weakened by many authors. In the present form, the theorem was proved in 
\cite{I,GI,Jo}. See \cite{Simon} for a detailed account.

A strong motivation to study such asymptotics first came in the end of 1940's
after Onsager's solution of the 2-dimensional Ising model and his observation that a 
2-spin correlation function in the model can be written as a Toeplitz determinant $D_n(f)$,
where $n$ denotes the distance between the spins.
For temperatures less than critical ($T<T_c$), the symbol of this Toeplitz determinant 
has an analytic logarithm in a neighborhood of the unit circle
and, moreover, $(\ln f)_0=0$. Therefore, Szeg\H o's theorem can be applied, and  one concludes that
$D_n(f)$ tends to a constant as $n\to\infty$.
Thus the correlation does not decay as the distance increases, which indicates the presence of
a long-range order, and hence, a magnetization. 
As $T\nearrow T_c$, however, 2 singularities of $f(z)$ approach the unit circle at $z=1$, and,
at $T=T_c$ merge into a single singularity {on} $C$; namely, a jump-type singularity at $z=1$
(see, e.g., \cite{MW}).
For $f(z)$ with such a singularity, the sum (\ref{Sf}) diverges, and Theorem \ref{szegothm}
can no longer be applied. In fact, it turns out \cite{MW} that in this case $D_n(f)$   
decays as $n^{-1/4}$ and, therefore, there exists no long-range order. 
For correlation functions arising in other situations, such as, e.g., the so-called emptiness formation probability
in the XY spin chain in a magnetic field \cite{DIZ,FA}, one obtains Toeplitz determinants
with both jump-type and root-type singularities, and in the most general situation
one is led to consider symbols of the form:
\be\la{fFH}
f(z)=e^{V(z)} z^{\sum_{j=0}^m \bt_j} 
\prod_{j=0}^m  |z-z_j|^{2\al_j}g_{z_j,\bt_j}(z)z_j^{-\bt_j},\qquad z=e^{i\th},\qquad
\theta\in[0,2\pi),
\ee
for some $m\ge 0$,
where
\begin{eqnarray}
&z_j=e^{i\th_j},\quad j=0,\dots,m,\qquad
0=\th_0<\th_1<\cdots<\th_m<2\pi;&\la{z}\\
&g_{z_j,\bt_j}(z)=
\begin{cases}
e^{i\pi\bt_j}& 0\le\arg z<\th_j\cr
e^{-i\pi\bt_j}& \th_j\le\arg z<2\pi
\end{cases},&\la{g}\\
&\Re\al_j>-1/2,\quad \bt_j\in\bbc,\quad j=0,\dots,m,&\la{cond-al}
\end{eqnarray}
and $V(e^{i\theta})$ is a sufficiently smooth function on the unit circle (see below)
with Fourier coefficients
\be\la{fourier}
V_k={1\over 2\pi}\int_0^{2\pi}V(e^{i\th})e^{-ki\th}d\th.
\ee
The canonical Wiener-Hopf factorization of $e^{V(z)}$ is given by 
\be\la{WienH}
e^{V(z)}=b_+(z) e^{V_0} b_-(z),\qquad b_+(z)=e^{\sum_{k=1}^\infty V_k z^k},
\qquad b_-(z)=e^{\sum_{k=-\infty}^{-1} V_k z^k}.
\ee

The condition (\ref{cond-al}) on $\al_j$ ensures the integrability of $f$.
Note that the size of the jump at $z_j$ is determined by the parameter $\bt_j$, and the root-type
singularity, by $\al_j$. We  assume that $z_j$, $j=1,\dots,m$, are genuine singular points, i.e.,
either $\al_j\neq 0$ or $\bt_j\neq 0$. However, the absence of a singularity
at $z=1$, i.e. the case $\al_0=\bt_0=0$, is allowed.  
Singularities of type (\ref{fFH}) are known
as Fisher-Hartwig singularities because of the work \cite{FH} where the authors
summarized a variety of applications of Toeplitz determinants with such symbols 
and presented a conjecture about the asymptotic form of $D_n(f)$ in this case. 
Due to the subsequent efforts of many workers, we have the following description of 
the asymptotics. 

Define the seminorm
\be
|||\bt|||=\max_{j,k}|\Re\bt_j-\Re\bt_k|,
\ee
where the indices $j,k=0$ are omitted if $z=1$ is not a singular point, i.e. if $\al_0=\bt_0=0$.
Note that in the case of a single singularity, we always have $|||\bt|||=0$. 

First, consider the situation when $|||\bt|||$ is strictly less then $1$. 
\begin{theorem}\la{asTop}
Let $f(z)$ be defined in (\ref{fFH}), $|||\bt|||<1$, $\Re\al_j>-1/2$, $\al_j\pm\bt_j\neq -1,-2,\dots$
for $j,k=0,1,\dots,m$, and
$V(z)$ satisfies the smoothness conditions
(\ref{Vcond}), (\ref{s1}) below. Then as $n\to\infty$,
\begin{multline}\la{asD}
D_n(f)=\exp\left[nV_0+\sum_{k=1}^\infty k V_k V_{-k}\right]
\prod_{j=0}^m b_+(z_j)^{-\al_j+\bt_j}b_-(z_j)^{-\al_j-\bt_j}\\
\times
n^{\sum_{j=0}^m(\al_j^2-\bt_j^2)}\prod_{0\le j<k\le m}
|z_j-z_k|^{2(\bt_j\bt_k-\al_j\al_k)}\left({z_k\over z_j e^{i\pi}}
\right)^{\al_j\bt_k-\al_k\bt_j}\\
\times
\prod_{j=0}^m\frac{G(1+\al_j+\bt_j) G(1+\al_j-\bt_j)}{G(1+2\al_j)}
\left(1+o(1)\right),
\end{multline}
where
$G(x)$ is Barnes' $G$-function \ci{Barnes}. The double product over $j<k$ is set to $1$
if $m=0$. The branches in (\ref{asD}) are determined as follows:
$b_\pm(z_j)^{-\al_j\pm\bt_j}=\exp\{(-\al_j\pm\bt_j)\sum_{k=1}^\infty V_{\pm k}z^{\pm k}\}$,
$(z_k z_j^{-1} e^{-i\pi})^{\al_j\bt_k-\al_k\bt_j}=\exp\{i(\th_k-\th_j-\pi)(\al_j\bt_k-\al_k\bt_j)\}$.
\end{theorem}

Note that since $G(-k)=0$, $k=0,1,\dots$,  formula (\ref{asD}) no longer 
represents the leading asymptotics if $\al_j+\bt_j$ or $\al_j-\bt_j$ is a negative 
integer for some $j$. Such degenerate cases can be handled by carrying the analysis to higher
order, but we present no further details here.

The smoothness condition on $V(z)$ assumed in Theorem \ref{asTop} is that 
\be\la{Vcond}
\sum_{k=-\infty}^\infty |k|^s |V_k|<\infty
\ee
holds for some $s$ such that
\be\la{s1}
s>
\frac{1+\sum_{j=0}^m\left[(\Im\al_j)^2+(\Re\bt_j)^2\right]}{1-|||\bt|||}.
\ee
Note that the condition $|||\bt|||<1$ is important here.

The Barnes' $G$-function first appeared in asymptotic Toeplitz theory in the work of Lenard \ci{Lenard}.
Theorem \ref{asTop} was proved by Widom \ci{W} in the case 
when $\Re\al_j>-1/2$, and all $\bt_j=0$,
and with a stronger condition on $V(z)$.
In \ci{B}, Basor extended the result to
$\Re\al_j>-1/2$, $\Re\bt_j=0$, and in \ci{Blocal}, to $\al_j=0$, $|\Re\bt_j|<1/2$.
In \ci{BS}, B\"ottcher and Silbermann established the result in the case that
$|\Re\al_j|<1/2$, $|\Re\bt_j|<1/2$. In \ci{Ehr}, Ehrhardt proved
the theorem for the full range of parameters, namely  
$\Re\al_j>-1/2$, $|||\bt|||<1$, and for $C^\infty$ functions $V(z)$.
These results were established by operator-theory methods
(see \ci{Ehr} for a review of these and other related results including an extension to
$\Re\al<-1/2$, $2\al\neq -1,-2,\dots$, when $f$ is replaced by a suitable distribution).
In \cite{DIKfh2}, the authors reprove the theorem by Riemann-Hilbert-Problem methods,
and relax the smoothness conditions on $V(z)$ to (\ref{Vcond}), (\ref{s1}).

Consider now the general case of Fisher-Hartwig symbols $f(z)$ with 
 the restriction $|||\bt|||< 1$ removed. Note first that $f(z)$ has several representations
of type (\ref{fFH}) with different sets of parameters $\bt_j$.
Namely, if each $\bt_j$ in (\ref{fFH}) such that either $\bt_j\neq 0$ or $\al_j\neq 0$  
is replaced by $\widehat \bt_j=\bt_j+n_j$, where $n_j$ are integers subject to the condition
$\sum_{j=0}^n n_j=0$, then the resulting function $f(z;n_0,\dots,n_m)$ is related to $f(z)$ 
in the following way:
\[
 f(z;n_0,\dots,n_m)=\prod_{j=0}^m z_j^{-n_j}f(z),
\]
i.e., it differs from $f(z)$ only by a constant. Each $f(z;n_0,\dots,n_m)$ so obtained
is called a FH-representation of the symbol. Denote by $\mathcal{M}$ the (finite) set of
FH-representations for which $\sum_{j=0}^m (\Re\widehat\bt_j)^2$ is minimal.
There exists a simple procedure (see \ci{DIKfh}) to solve this discrete variational problem and 
to construct $\mathcal{M}$ explicitly.
One can show that there is always a FH-representation with $|||\widehat\bt|||\le 1$, and we have 
the following 2 mutually exclusive possibilities:
\begin{itemize}
\item
If there exists a FH-representation such that $|||\widehat\bt|||<1$ then it turns out that
this FH-representation is the {\it single} element of $\mathcal{M}$.
In particular, if $|||\bt|||<1$, the set  $\mathcal{M}$ consists of a
single element corresponding to all $n_j=0$, and Theorem \ref{BT} below reduces to
Theorem \ref{asTop}. 
\item
If there exists a FH-representation such that $|||\widehat\bt|||=1$ then $\mathcal{M}$ consists
of several (at least 2) elements. 
\end{itemize}

The set $\mathcal{M}$ is called non-degenerate if it contains no representations
for which $\al_j+\widehat\bt_j$ or $\al_j-\widehat\bt_j$ is a negative 
integer for some $j$. The general result is as follows.

\begin{theorem}\la{BT}
Let $f(z)$ be given in (\ref{fFH}), $V(z)$ satisfy the condition (\ref{Vcond}) above 
for some sufficiently large $s$ (depending only on $\al_j$, $\bt_j$), and
$\Re\al_j>-1/2$, $\bt_j\in\bbc$, $j=0,1,\dots,m$.
Let $\mathcal{M}$ be non-degenerate.
Then, as $n\to\infty$,
\be\la{asDgen}
D_n(f)=\sum\left[\left(\prod_{j=0}^m z_j^{n_j}\right)^n 
\mathcal{R}(f(z;n_0,\dots,n_m))(1+o(1))\right],
\ee
where the sum is over all FH-representations in $\mathcal{M}$.
Each $\mathcal{R}(f(z;n_0,\dots,n_m))$ stands for the right-hand side 
of the formula (\ref{asD}), without the error term,
corresponding to $f(z;n_0,\dots,n_m)$.
\end{theorem}

An explicit lower bound on $s$ (depending on $\bt_j$, $\al_j$) similar to (\ref{s1})
is given in \ci{DIKfh}.

This theorem was conjectured by Basor and Tracy \ci{BT} and proved in \ci{DIKfh}.

Hankel and Toeplitz+Hankel determinants are also of interest.
Let $w(x)$ be an integrable function on a subset $J$ of $\bbr$. 
Then the Hankel determinant with symbol $w(x)$ supported on $J$ is given by 
\be
D_n^H(w(x))=\det\left(\int_J x^{j+k}
  w(x)dx\right)_{j,k=0}^{n-1}.
\ee
When $J$ is a finite interval -- we then set $J=[-1,1]$ without loss of generality --
Hankel determinants are related to Toeplitz determinants by the following formulae \ci{DIKfh}, 
involving the orthogonal polynomials (\ref{op}):
\be\la{wf}
w(x)={f(e^{i\th})\over|\sin\th|},\qquad x=\cos\th,\quad x\in[-1,1];
\ee
\be\la{THconnection}
[D_n^H(w(x))]^2={\pi^{2n}\over 4^{(n-1)^2}}
{(\chi_{2n}+\phi_{2n}(0))^2\over\phi_{2n}(1)\phi_{2n}(-1)}D_{2n}(f(z)).
\ee

A particularly interesting class of Toeplitz+Hankel determinants appearing in the theory of classical groups and 
its applications to random matrices and statistical mechanics
(see, e.g., \ci{BR,FF,KM}) is defined as follows for even $f(e^{i\th})=f(e^{-i\th})$ (for even $f$
the matrices involved are symmetric): 
\be\la{intT+H}
\det (f_{j-k}+f_{j+k})_{j,k=0}^{n-1},\quad
\det (f_{j-k}-f_{j+k+2})_{j,k=0}^{n-1},\quad
\det (f_{j-k}\pm f_{j+k+1})_{j,k=0}^{n-1}.
\ee
They are related to Hankel determinants with symbols on $[-1,1]$
by the expressions
\begin{eqnarray}
\det (f_{j-k}+f_{j+k})_{j,k=0}^{n-1}={2^{n^2-2n+2}\over\pi^n} 
D_n^H(f(e^{i\theta(x)})/\sqrt{1-x^2}),\\
\det (f_{j-k}-f_{j+k+2})_{j,k=0}^{n-1}={2^{n^2}\over\pi^n} 
D_n^H(f(e^{i\theta(x)})\sqrt{1-x^2}),\la{TH-}\\
\det (f_{j-k}+f_{j+k+1})_{j,k=0}^{n-1}={2^{n^2-n}\over\pi^n} 
D_n^H(f(e^{i\theta(x)})\sqrt{1+x\over 1-x}),\la{TH+1}\\
\det (f_{j-k}-f_{j+k+1})_{j,k=0}^{n-1}={2^{n^2-n}\over\pi^n} 
D_n^H(f(e^{i\theta(x)})\sqrt{1-x\over 1+x}).\la{TH-1}
\end{eqnarray}

Asymptotic formulae for Hankel and Toeplitz+Hankel determinants
with Fisher-Hartwig singularities, whose derivation
was based on the above theorems for Toeplitz determinants, an asymptotic 
Riemann-Hilbert-Problem analysis of
the polynomials (\ref{op}), and the above relations, are presented in \ci{DIKfh}.
For other asymptotic results, see \ci{BEnonsym,BE4,BR2,BCW,CI}.  

For related asymptotic results on an important class (see Section 4) of Toeplitz determinants
when the symbol is supported on an arc of the unit circle, see \ci{Warc,K_arc,K,DKV}.

Theorem \ref{BT} and asymptotic formulae for Hankel and Toeplitz+Hankel determinants
find applications, e.g., for correlation functions in the $XY$ spin chain in a magnetic field 
mentioned above, in the theory of the impenetrable Bose gas \ci{FF,Ovchinnikov}, in random matrix
conjectures for average values of
Riemann's $\zeta$-function, and Dirichlet's $L$-functions \ci{KS,GHK,BK}.

For a more detailed discussion of the material presented in this section so far, the reader is
referred to \ci{DIKfh}.

A related area of interest is the asymptotic analysis of Hankel determinants whose
symbol has Fisher-Hartwig singularities and is supported on the whole real line, or the half-line.
In particular, in the Gaussian Unitary Ensemble of random matrix theory, 
the correlation function of products of powers of the absolute values of the characteristic
polynomial is precisely such a Hankel determinant $D_n^H(w)$: namely, the symbol is supported on $\bbr$
and given by $w(x)=\exp(-x^2)\prod_{j=1}^m |x-\mu_j|^{2\al_j}$, $\mu_j\in\bbr$, $\Re\al_j>-1/2$.
This determinant is also related to the 1-dimensional impenetrable Bose gas and
conjectures for mean values of Riemann's $\zeta$-function on the critical line. 
For a discussion of the results in this area, see \ci{Kduke,IK,IKO}.
For analysis of some other Hankel determinants appearing in random matrix models,
see \cite{BI,EM}. For a recent application of Hankel determinants in the six-vertex model
see \ci{Iz,BL1,BL2,BL3,BF}. 

Note that the importance of Fisher-Hartwig singularities appears to stem from the following feature.
The asymptotics of the orthogonal polynomials at the location of such a singularity
are described by the confluent hypergeometric function \ci{DIKfh}. The two independent parameters of such
functions are related to the parameters $\al$ and $\bt$ of the singularity (for $\bt=0$
confluent hypergeometric functions reduce to Bessel functions). The location of the 
singularity corresponds to the {\it single} finite branch point of the {\it confluent} hypergeometric
functions. As hypergeometric functions (which depend on 3 parameters) have 
2 finite branch points, 
we would not be able to confine ourselves to hypergeometric functions if we wanted
to consider a singular point which generalizes Fisher-Hartwig in some essential way. 
Roughly speaking, a Fisher-Hartwig singularity is the most general hypergeometric singular point.
Modifications, of course, are possible: e.g., the end points of the interval $[-1,1]$ can be
regarded as modified Fisher-Hartwig singularities for a Hankel determinant with symbol on
$[-1,1]$ as discussed above.

\section{Transition asymptotics}
A natural question to ask is how the transition between various asymptotic regimes of 
the previous section occurs. Consider once again the 2-spin correlation function for the 
2-dimensional Ising model discussed above, which is a Toeplitz determinant.
As $T\nearrow T_c$ a transition between the Szeg\H o asymptotics and the Fisher-Hartwig 
asymptotics takes place. It was first investigated in \ci{WMTB,MTW,T}, and the authors found that
if $T\to T_c$ and $n\to\infty$ in such a way that $x\equiv (T_c-T)n$ is fixed,
then the determinant is given in terms of Painlev\'e III (reducible to Painlev\'e V) functions. 
This transition corresponds
to the emergence of one Fisher-Hartwig singularity with $\al=0$, $\bt=-1/2$ at $z_0=1$.
The condition that  $x\equiv (T_c-T)n$ is fixed was removed in \ci{CIK} where uniform asymptotics
were obtained for any $\al$, $\Re\al>-1/2$, $\bt\in\bbc$ in terms of Painlev\'e V functions.
Namely, consider the following symbol
\begin{equation}\label{symbol}
f_t(z)=(z-e^{t})^{\alpha+\beta}(z-e^{-t})^{\alpha-\beta}z^{-\alpha+\beta}
e^{-i\pi(\alpha+\beta)}e^{V(z)},\qquad \al\pm\bt\neq -1,-2,...
\end{equation}
where $t\geq 0$ is sufficiently small (in the above example of the Ising model, 
$t=\mathrm{const} (T_c-T)$),
$V(z)$ is analytic in a neighborhood of $C$, and $\alpha, \beta\in\mathbb C$ with 
$\Re\alpha>-\frac{1}{2}$. The singularities of the symbol
are at the points $e^{\pm t}$.
If $t=0$ the symbol possesses a Fisher-Hartwig singularity
at $z=0$ and Theorem \ref{asTop} applies to $D_n(f_0)$. If $t>0$ then  $f_t(z)$ is analytic 
in a neighborhood of $C$, and Szeg\H o's Theorem \ref{szegothm} applies. We have \ci{CIK}
\begin{theorem}\label{es}
Let $\alpha, \beta\in\mathbb C$ with
$\Re\alpha>-\frac{1}{2}$ and let $s_\de$ denote a sector  $-\pi/2+\de<\arg x<\pi/2-\de$, 
$0<\de<\pi/2$.
Let $f_t$ be given by (\ref{symbol}) and consider the Toeplitz determinants $D_n(f_t)$
defined by (\ref{TD}) corresponding to this symbol.
There exists a finite set $\{x_1, \ldots , x_k\}\in s_\de$ (with $k=k(\alpha,\beta)$ and
$x_j=x_j(\alpha,\beta)\neq 0$) such that 
there holds the following expansion as $n\to\infty$
with the error term uniform for $0<t<t_0$ (with $t_0$ sufficiently small)
as long as $2nt$ remains bounded away from the set $\{x_1, \ldots , x_k\}$:
\begin{multline}\label{expansion Dn}
\ln D_n(f_t)=n V_0+(\al+\bt)nt+
\sum_{k=1}^{\infty}k\left[V_k-(\alpha+\beta)\frac{e^{-tk}}{k}\right]
\left[V_{-k}-(\alpha-\beta)\frac{e^{-tk}}{k}\right]\\
+\ln\frac{G(1+\alpha+\beta)G(1+\alpha-\beta)}{G(1+2\alpha)}
+\Om(2nt)+o(1),
\end{multline}
where $G(z)$ is Barnes' G-function, and
\begin{equation}
\label{def Omega}
\Om(2nt)=\int_0^{2nt}\frac{\sigma(x)-\alpha^2+\beta^2}{x}dx+(\alpha^2-\beta^2)\ln
2nt.
\end{equation}
The function $\sigma(x)$ is a particular solution to the 
Jimbo-Miwa-Okamoto $\sigma$-form \cite{J, jm2} of the
Painlev\'e V equation
\be
\left(x\frac{d^2\sigma}{dx^2}\right)^2 = \left(\sigma
-x\frac{d\sigma}{dx} +2\left(\frac{d\sigma}{dx}\right)^2 +
2\alpha\frac{d\sigma}{dx}\right)^2
-4\left(\frac{d\sigma}{dx}\right)^2\left(\frac{d\sigma}{dx} +\alpha
+\beta\right) \left(\frac{d\sigma}{dx} +\alpha -\beta\right).\label{sigma5}
\ee
This solution has the
following asymptotics for $x>0$: \be\label{westintro}
\sigma(x)=\begin{cases}\al^2-\bt^2+
\frac{\alpha^2-\beta^2}{2\alpha}\{x-x^{1+2\al}C(\alpha,
\beta)\}(1+O(x)),& x\to 0,\quad 2\al\notin\mathbb Z\cr
\al^2-\bt^2+O(x)+O(x^{1+2\al})+O(x^{1+2\al}\ln x),& x\to
0,\quad 2\al\in\mathbb Z\cr x^{-1+2\alpha}e^{-x}
\frac{-1}{\Gamma(\alpha-\beta)\Gamma(\alpha+\beta)} \left(1 +
O\left(\frac{1}{x}\right)\right),& x\to +\infty,
\end{cases}
\ee
with
\be\label{Cab}
C(\al,\bt)=
\frac{\Gamma(1+\al+\bt)\Gamma(1+\al-\bt)}{\Gamma(1-\al+\bt)\Gamma(1-\al-\bt)}
\frac{\Gamma(1-2\al)}{\Gamma(1+2\al)^2}\frac{1}{1+2\al},
\ee
where $\Gamma(z)$ is Euler's $\Gamma$-function.
The path of integration in (\ref{def Omega})
is such as to avoid the set $\{x_1, \ldots , x_k\}$
and is contained within the sector $s_\de$.
\end{theorem}

Note that (\ref{sigma5}) is the $\si$-form of the Painlev\'e V equation
\begin{equation}\label{PVintro}
u_{xx}=\left(\frac{1}{2u}+\frac{1}{u-1}\right)u_{x}^2-\frac{1}{x}u_x+
\frac{(u-1)^2}{x^2}\left(A u+\frac{B}{u}\right)+\frac{C u}{x}
+D\frac{u(u+1)}{u-1},
\end{equation}
with the parameters $A, B, C, D$ given by
\begin{equation}\label{ABCD}
A=\frac{1}{2}(\alpha-\beta)^2,\qquad B=-\frac{1}{2}(\alpha+\beta)^2,\qquad
C=1+2\beta, \qquad D=-\frac{1}{2}.
\end{equation}

The points $x_j$ refer to possible poles of $\si(x)$.
In the case when $\al$ is real and $\bt$ is purely imaginary, one can show \ci{CIK}
that $\si(x)$ is real analytic for $x>0$, 
and the path of integration can therefore be chosen along the real axis.

If one takes the limit as $t \to 0$ on the r.h.s. of (\ref{expansion Dn}),
one obtains the correct form of the appropriate Fisher-Hartwig
asymptotics as given by Theorem \ref{asTop}.

Since the asymptotics of $D_n(f)$ are known both at $t=0$ and $t=t_0$,
one obtains an amusing identity for the Painlev\'e function $\si(x)$:
\be
\Om(+\infty)=-\ln\frac{G(1+\alpha+\beta)G(1+\alpha-\beta)}{G(1+2\alpha)}.
\ee

Methods used in \ci{CIK} to prove Theorem \ref{es} can be adapted to describe other transition
regimes, e.g., two singularities approaching each other along the unit circle, or
emergence of an arc on which $f(z)=0$.
These situations arise for other correlation functions in integrable models \cite{FA} and
appear in the application of random matrix theory to the theory of $L$-functions \cite{BK}.

\section{Asymptotics for Fredholm determinants}
We now consider another type of double-scaling limit for Toeplitz determinants which yields
interesting Fredholm determinants and, after certain analysis, allows us to obtain asymptotics 
of the latter. Note that this approach combined with Riemann-Hilbert-Problem techniques
allows us to obtain the full asymptotics of these
Fredholm determinants including the multiplicative constants which resisted other methods:
see \ci{Kgaprev} for a short review of the approach and \ci{K,DIKZ,DIKairy,BBD,DKV} for details.
For a review of other applications of Riemann-Hilbert problems to Toeplitz and Fredholm
determinants see \ci{D}. For analysis of some other Fredholm determinants, see \ci{W4,KKMST,CIKp2} and
references in Introduction.

Let $f(e^{i\th};n)=1$ on the arc $2s/n\le\th\le2\pi-2s/n$, $0<s<n$, and $f(z;n)=0$ on the rest of 
the unit circle.
Then the Fourier coefficients are $f_0=1-2s/(n\pi)$, $f_j=-\frac{\sin(2sj/n)}{\pi j}$, $j\neq 0$.
In the limit of growing $n$ and, accordingly, a closing arc,
\be\la{Tsine}
\lim_{n\to\infty}D_n(f(z;n))=\det(I-K_{sine}^{(s)}),
\ee
where $K_{sine}^{(s)}$ is the trace-class operator on $L^2(-s,s)$ with kernel
\be\la{sine}
K_{sine}(x,y)=\frac{\sin(x-y)}{\pi(x-y)}.
\ee
In the Gaussian Unitary Ensemble of random matrix theory (and many other random matrix ensembles
\cite{DGbook,Schlein}), the Fredholm sine-kernel determinant $\det(I-K_{sine}^{(s)})$
describes, in the bulk scaling limit, the probability that an interval of length $2s$
contains no eigenvalues. 
Of interest are the asymptotics of $\det(I-K_{sine}^{(s)})$ when $s$ is large.

\begin{theorem}\la{Fsinethm} 
Let $K_{sine}^{(s)}$ be the operator acting on $L^2(-s,s)$, $s>0$, with kernel (\ref{sine}). Then 
as $s\to+\infty$,
\be\la{dyson}
\det(I-K_{sine}^{(s)})=c_{sine}s^{-{1\over 4}}\exp\left(-{s^2\over 2}\right)\left[
1+O(s^{-1})\right],
\qquad  c_{sine}=2^{1/12}e^{3\zeta'(-1)},
\ee
and $\zeta'(x)$ is the derivative of Riemann's zeta function.
\end{theorem}

We note that $s(d/ds) \ln \det(I-K_{sine}^{(s)})$
satisfies \ci{JMMS,DIZ} a form of the Painlev\'e V equation.
In particular, this fact enables one to reconstruct the full asymptotic series of the 
logarithmic derivative in the inverse powers of $s$
from the first few terms provided the existence of such asymptotic expansion is established.
However, the multiplicative constant $c_{sine}$ is not determined this way.

Theorem \ref{Fsinethm} was conjectured by Dyson \ci{Dyson} who used, in particular, (\ref{Tsine})
and an earlier result of Widom on Toeplitz determinants with a symbol which vanishes on a fixed arc of 
the unit circle \ci{Warc}.
The leading asymptotic term was proved by Widom \ci{Wsine}, and the lower-order terms apart from 
$c_{sine}$, in other words the expansion for the derivative $(d/ds) \ln \det(I-K_{sine}^{(s)})$, by
Deift, Its, and Zhou \ci{DIZ} using Riemann-Hilbert methods. 
Application of a more detailed Riemann-Hilbert analysis to Toeplitz determinants 
allowed the authors in \ci{K,DIKZ} to extend the result of Widom \ci{Warc} to
varying arcs, and the relation (\ref{Tsine}) then produced 
the asymptotics of $\det(I-K_{sine}^{(s)})$ {\it including} $c_{sine}$,
which completed the proof of Theorem \ref{Fsinethm}. 
An alternative proof of the theorem was given independently by Ehrhardt \ci{Esine} who used
(different) methods of operator theory.

Recall that $f(z;n)$ was defined above on the arc whose end-points converge to $z=1$ as $n\to\infty$.
We now modify the definition of $f(z;n)$ by placing a Fisher-Hartwig singularity at $z=1$.
Namely, consider the symbol  
$F(z;n)=|z-1|^{2\alpha}z^{\bt}e^{-i\pi\bt}$, $z=e^{i\th}$, 
on the arc $2s/n\le\th\le2\pi-2s/n$, $0<s<n$, 
and $F(z;n)=0$ on the rest of the unit circle.
We then obtain \ci{DKV}
\be\la{Tch}
\lim\limits_{n\to \infty}\frac{D_n(F(z;n))}{D_n(F(z;\infty))}=\det(I-K^{(\al,\bt,s)}_{ch}),
\ee
where $K^{(\al,\bt,s)}_{ch}$ is the trace-class operator on $L^2(-s,s)$ with kernel 
\be
\label{ch}
K^{(\alpha,\beta,s)}_{ch}(u,v)=\frac 1{2\pi i}\frac{\Gamma(1+\alpha+\beta)
\Gamma(1+\alpha-\beta)}{\Gamma(1+2\alpha)^2}
\frac{A(u)B(v)-A(v)B(u)}{u-v},
\ee
where
\begin{eqnarray}
&A(x)=g_\beta^{1/2}(x)|2x|^{\alpha}e^{-ix}\phi(1+\alpha+\beta,1+2\alpha,2ix),&\nonumber\\
&B(x)=g_\beta^{1/2}(x)|2x|^{\alpha}e^{ix}\phi(1+\alpha-\beta,1+2\alpha,-2ix),&\nonumber\\
&g_\beta(x)=\begin{cases}
e^{-\pi i\beta},& x>0, \cr
e^{\pi i\beta}, & x<0. 
\end{cases},\qquad \alpha,\beta\in\bbc,\quad \Re{\alpha}>-1/2,\quad \al\pm\bt\neq-1,-2,\dots&\nonumber
\end{eqnarray}
Here $\phi(a,c,z)$ is the confluent hypergeometric function (see, e.g., \cite{Abr})
\begin{equation}\label{phidef}
\phi(a,c,z) = 1+\sum_{n=1}^{\infty}\frac{a(a+1)\cdots(a+n-1)}{c(c+1)\cdots(c+n-1)}\frac{z^n}{n!}.
\end{equation}
The kernel $K^{(\al,\bt,s)}_{ch}$ appears in the representation theory of the infinite-dimensional
unitary group \ci{BO,BD}, and the logarithmic derivative $(d/ds) \ln \det(I-K^{(\al,\bt,s)}_{ch})$
is related to a solution of the Painlev\'e V equation. If we set $\al=\bt=0$, the kernel reduces
to the sine-kernel (\ref{sine}).

\begin{theorem}\la{Fchthm}
Let $K^{(\al,\bt,s)}_{ch}$ be the operator acting on 
$L^2(-s,s)$, $s>0$, with kernel (\ref{ch}). Then as $s\to+\infty$,
\be
\label{Ps}
\det(I-K^{(\al,\bt,s)}_{ch})= 
\frac{\sqrt{\pi}G^2(1/2)G(1+2\alpha)}{2^{2\alpha^2}G(1+\alpha+\beta)G(1+\alpha-\beta)}
s^{-\frac 14-\alpha^2+\beta^2}
\exp\left(-\frac{s^2}2+2\alpha s\right)\left[1+O(s^{-1})\right],
 \ee
where $G(x)$ is Barnes' G-function.
\end{theorem}

This theorem was proved in \ci{DKV} using the relation (\ref{Tch}) and a Riemann-Hilbert 
analysis. The theorem reduces to Theorem \ref{Fsinethm} if $\al=\bt=0$
(recall that $2\ln G(1/2)=(1/12)\ln2-\ln\sqrt{\pi}+3\zeta'(-1)$). 

A particular case of the determinant $D_n(F(z;n))$ with $\bt=0$ is related, via a Hankel determinant
and the formula (\ref{THconnection}),
to the following Bessel-kernel determinant $\det(I-K_{Bessel}^{(a,s)})$ 
on $(0,s)$, where the kernel
\be\la{Bkernel}
K_{Bessel}^{(a,s)}(x,y)=\frac{\sqrt{y}J_{a}(\sqrt{x})J'_a(\sqrt{y})-
\sqrt{x}J_{a}(\sqrt{y})J'_{a}(\sqrt{x})}{2(x-y)},
\ee
and $J_a(x)$ is Bessel function.
In the Jacobi Unitary Ensemble of random matrix theory
(and many other ensembles with a so-called hard edge), the Bessel-kernel determinant 
$\det(I-K_{Bessel}^{(a,s)})$
describes, in the (left) edge scaling limit, the probability that the interval $(0,s)$
contains no eigenvalues. In other words, it describes the distribution of the 
extreme (smallest) eigenvalue.

\begin{theorem}\la{detbessel}
Let $K_{Bessel}^{(a,s)}$ be the operator acting on $L^2(0,s)$, $s>0$, with kernel (\ref{Bkernel})
where $\Re a> -1$. 
Then as $s\to+\infty$,
\be
\label{widom}
\det(I-K_{Bessel}^{(a,s)})=
c_{Bessel}(a) s^{-a^2/4}\exp\left(-{s\over 4}+a\sqrt{s}\right)\left[1+O(s^{-1/2})\right],
\qquad c_{Bessel}(a) =\frac{G(1+a)}{(2\pi)^{a/2}}.
\ee
\end{theorem}
These asymptotics were conjectured by Tracy and Widom \ci{TWbessel} and proved in \ci{DKV}.
An alternative proof of
the particular case $|\Re a|<1$ is given in \ci{Ebessel} by methods of operator theory.

Finally, we consider the case of the so-called Airy kernel.
Let $w(x)=e^{-4xn}$ be supported on $J=[0,1+s(2n)^{-2/3}]$ for a fixed $s\in\bbr$,
and let $D_n^H(w)$ be the corresponding
Hankel determinant. Then
\be\la{Hairy}
\lim_{n \to \infty}D_{n}^H\left(1 + \frac{s}{(2n)^{2/3}}\right)
=\det\left(I-K_{Airy}^{(s)}\right),
\ee
where $K_{Airy}^{(s)}$ is the trace-class operator on $L^2(s,+\infty)$ with kernel 
\be\la{airy}
K_{Airy}^{(s)}(x,y) ={\Ai(x)\Ai'(y)-\Ai(y)\Ai'(x)\over x-y}.
\ee
Here $\Ai(x)$ is the Airy function (see, e.g., \ci{Abr}).

In the Gaussian Unitary Ensemble of random matrix theory
(and many other ensembles with a so-called soft edge), 
the Airy-kernel determinant $\det(I-K_{Airy}^{(s)})$
describes, in the (right) edge scaling limit, the probability that the interval $(s,+\infty)$
contains no eigenvalues. In other words, it describes the distribution of the 
extreme (largest) eigenvalue.

\begin{theorem}\la{Fairy}
Let $K_{Airy}^{(s)}$ be the operator acting on $L^2(s,+\infty)$, $s\in\bbr$,
with kernel (\ref{airy}). Then as $s\to -\infty$,
\be
F_{TW}(s)\equiv\det(I-K_{Airy}^{(s)})=c_{Airy}|s|^{-1/8}\exp\left(-{|s|^3\over 12}\right)
\left[1+O(|s|^{-3/2})\right],\qquad c_{Airy}=2^{1/24}e^{\ze'(-1)},
\ee
\end{theorem}

The distribution $F_{TW}(s)$ is known as the Tracy-Widom distribution.
Tracy and Widom showed that \ci{TW} 
\begin{equation}\label{TW}
F_{TW}(s) = \exp \left \{ -\int_{s}^{\infty}(x-s)u^{2}(x)dx\right \},
\end{equation}
where $u(x)$ is the Hastings-McLeod solution of the Painlev\'e II
equation
\begin{equation}\label{ds:30}
u''(x)=xu(x)+2u^3(x)\,,
\end{equation}
specified by the following asymptotic condition:
\begin{equation}\label{ds:40}
u(x)\sim \Ai(x)\qquad \mbox{as}\quad x\to+\infty.
\end{equation}
The asymptotics of the logarithmic derivative $(d/ds)\ln F_{TW}(s)$ follow,
up to a constant (which is in fact zero), from (\ref{ds:40})
and the known asymptotics of the Hastings-McLeod solution at $-\infty$. 
The constant $c_{airy}$ (as well as $c_{Bessel}$ above) was conjectured by Tracy and Widom
using numerical computations and an analogy with the Dyson formula (\ref{dyson}).
The full proof of Theorem \ref{Fairy} was given in \ci{DIKairy} using (\ref{Hairy}). 

The determinant $\det(I-K_{Airy}^{(s)})$ also describes 
the distribution of the longest increasing 
subsequence of random permutations. Namely, let $\pi=i_1 i_2\cdots i_N$
be a permutation in the group $S_N$ of permutations of $1,2,\dots,N$. Then a subsequence
$i_{k_1},i_{k_2},\dots i_{k_r}$, $k_1<k_2<\cdots<k_r$,  
of $\pi$ is called an increasing subsequence of length
$r$ if $i_{k_1}<i_{k_2}<\cdots<i_{k_r}$. Let $\ell_N(\pi)$ denote the length of a longest
increasing subsequence of $\pi$ and let $S_N$ have the uniform probability distribution.
Then $\ell_N(\pi)$ is a random variable, and 
\be\la{prob2}
F_{TW}(s)=\lim_{N\to\infty}\mbox{Prob }\{\pi\in S_N: 
(l_N(\pi)-2\sqrt{N})N^{-1/6}\le s\}
\ee
This result was obtained by Baik, Deift, and Johansson \ci{BDJ} from
the double-scaling limit $N\to\infty$, $n\le N\sim\lb$, of the 
Toeplitz determinant $\De_{n,\lb}=D_n(\exp\{\sqrt{\lb}(z+z^{-1})\})$.
As shown earlier by Gessel \ci{Ges}, this determinant is precisely the following generating function:
\be
\De_{n,\lb}=\sum_{N=0}^\infty u_n(N)\frac{\lb^N}{N!^2},\qquad
u_n(N)=\#(\mbox{permutations $\pi$ in $S_N$ with $\ell_N(\pi)\le n$}).
\ee
An alternative proof of Theorem \ref{Fairy} based on this determinant 
was given in \ci{BBD}.

By the Robinson-Schensted-Knuth correspondence (see, e.g., \ci{Andrews})
a permutation $\pi$ is related to a pair of Young tableaux (of the same shape) 
of integer plane partitions of $N$. 
The number $l_N(\pi)$ is the length of the first row of the related tableaux. 

We also note that random permutations are related to last passage percolation and
random vicious walks \ci{F,B1,B2}. 

There are many related results and extensions of the above results on
random partitions and permutations, which, in particular, involve
asymptotic analysis of special Toeplitz, Hankel, and Toeplitz+Hankel determinants. 
This large and growing research area has many connections 
to geometry, group representation theory, and integrable models. 
For details and a selection of results, 
see \ci{BDJ,J1,J2,BOO,O,BR,BR2,ITW,FS} and references therein.

\section*{Acknowledgement}
I thank Florian Sobieczky for inviting me to the Alp-workshop 2009. I am also grateful to
Percy Deift for useful comments.

\end{document}